\documentclass{article}
\usepackage{graphicx}
\begin{document}

\title{Quantum information via state partitions and the context translation principle}
\author{Karl Svozil\\
Institut f\"ur Theoretische Physik, University of Technology Vienna,  \\
Wiedner Hauptstra\ss e 8-10/136, A-1040 Vienna, Austria\\
email: svozil@tuwien.ac.at, homepage: http://tph.tuwien.ac.at/~svozil}

\maketitle

\begin{abstract}
For many--particle systems, quantum information in base $n$
can be defined by partitioning the set of states according to the outcomes
of $n$--ary (joint) observables.
Thereby, $k$ particles can carry $k$ nits.
With regards to the randomness of single outcomes,
a context translation principle is proposed.
Quantum randomness is related
to the uncontrollable degrees of freedom of the measurement interface,
thereby translating a mismatch between the state prepared and the state measured.
\end{abstract}

%\pacs{03.67.-a,03.67.Mn,03.65.Ta}
%\keywords{Quantum information, foundations of quantum mechanics, measurement theory}

\section{Information in many--particle quantum systems}

The preparation of a single particle $n$--state quantum system in a single state constitutes
the operationalization of a {\em nit,} or {\em qunit.}
Likewise, the occurrence of an outcome of an observable with $n$ possible outcomes
can be associated with accessing a  nit of information.
For a single particle observable,
this is associated with choosing a vector from an orthogonal basis of $n$--dimensional Hilbert space.
In the many--particle case,
nits may not only be localized at single particle observables, since
due to entanglement, the nits may be distributed
over the particles by representing joint particle properties.

In what follows we shall review
and extend formal generalizations \cite{DonSvo01,svozil-2002-statepart-prl}
of the single particle two--state case
to an arbitrary finite number of particles
with an arbitrary finite number of different measurement outcomes per particle.
Thereby, we define a nit as a radix $n$ measure of quantum information which
is based on state partitions associated with the outcomes of
$n$--ary observables.
We shall demonstrate the following property:
{\em
$k$ particles specify $k$ nits in such a way that $k$
measurements of comeasurable observables with $n$ possible outcomes are necessary to determine the information.}
Stated pointedly,
{\em
$k$ particles can carry $k$ nits.
}

Conceptually, such properties have been previously proposed
by Zeilinger \cite{zeil-99} as a
foundational principle for quantum mechanics.
Zeilinger merely considered two--state systems of two and three particles,
yet an informal hint for higher--dimensional single quantum systems
is in  footnote 6 of \cite[p. 635]{zeil-99}.
There is a slight difference in the approach of Zeilinger and the author:
whereas here the logico--algebraic properties are studied `top-down' by assuming
Hilbert space quantum mechanics and arriving at the foundational principle
purely deductively,
Zeilinger and Brukner \cite{zeil-bruk-02}
reconstruct certain features of quantum physics by treating this  principle `bottom--up'
as an axiom.

\subsection{Definition}

For a single $n$--state particle, the nit can be formalized as a fine-grained partition
of $n$ orthogonal states; i.e.,
if the set of orthogonal states is represented by $\{1,\ldots ,n\}$,
then the nit is defined by choosing one element of the set
$
%\begin{equation}
\{\{1\} ,\ldots ,\{n\}\}
%\label{2002-statepart-snit}
%\end{equation}
$.

The generalization to $k$ particles involves the construction of
$k$ partitions of the product states with
$n$ elements per partition in such a way that
every single product state is obtained by the set theoretic intersection of
$k$ elements of all the different partitions.
That is, the partitions which properly represent the set of nits
have to be defined to obey the following properties:
(i) every set theoretic intersection of single elements of the $k$ partitions,
one element per partition, yields a single product state,
and
(ii) the union of all these
intersections obtained by (i) is just the set of product states.
Every single such partition can be interpreted as a nit.
For their implementation, we shall adopt an $n$--ary search strategy.

%---->>> Remark Toffoli-coworker

In the following, the standard orthonormal basis of
$n^k$--dimensional Hilbert space is identified with the set of states $S=\{1,2,\ldots ,n^k\}$; i.e.,
(superscript `$T$' indicates transposition)
%\begin{equation}
%\begin{array}{llll}
%1 &\equiv& (1,\ldots,0)^T\equiv \mid 1_1,\ldots ,1_k\rangle = \mid 1_1\rangle \otimes \cdots \otimes \mid 1_k\rangle ,\\
%  &\vdots&\\
%n^k &\equiv& (0,\ldots,1)^T\equiv \mid n_k,\ldots ,n_k\rangle = \mid n_k\rangle \otimes \cdots \otimes \mid n_k\rangle .\\
%\end{array}
%\label{2002-statepart-psma}
%\end{equation}
$
1 \equiv (1,\ldots,0)^T\equiv \mid 1_1,\ldots ,1_k\rangle = \mid 1_1\rangle \otimes \cdots \otimes \mid 1_k\rangle
$,
$\ldots $,
$
n^k \equiv (0,\ldots,1)^T\equiv \mid n_1,\ldots ,n_k\rangle
$.
Here, the single particle states are labelled by $1_1$ through $n_k$, respectively.
Tensor product states are formed and ordered lexicographically ($0<1$).

The nit operators are defined via diagonal matrices
which contain $n^{k-1}$ equal amounts of $n$ mutually different numbers
such as different primes $q_1,\ldots ,q_n$; i.e.,
\begin{equation}
\begin{array}{llll}
F_1&=& {\rm diag} (\underbrace{\underbrace{q_1,\ldots ,q_1}_{n^{k-1}\;{\rm times}},\ldots ,\underbrace{q_n,\ldots ,q_n}_{n^{k-1}\;{\rm times}}}_{n^0=1\;{\rm times}}),\\
F_2&=& {\rm diag} (\underbrace{\underbrace{q_1,\ldots ,q_1}_{n^{k-2}\;{\rm times}},\ldots ,\underbrace{q_n,\ldots ,q_n}_{n^{k-2}\;{\rm times}}}_{n^1\;{\rm times}}),\\
  &\vdots&\\
F_k&=& {\rm diag} (\underbrace{q_1,\ldots ,q_n}_{n^{k-1}\;{\rm times}}).
\end{array}
\label{2002-statepart-nitopgen}
\end{equation}
`${\rm diag}(a,b,\ldots )$' stands for the diagonal matrix with $a,b,\ldots $
at the diagonal entries.
The operators implement an $n$--ary search filter,
separating the search space into $n$ equal partitions of states,
such that successive applications of all such filters
renders a single state.
In this simplest, nonentangled, case,
the meaning of the $i$'th filter or nit operator $F_i$, $1\le i\le k$,
can be expressed as the proposition,
{\em `the $i$'th particle is in state $q_1, \ldots ,q_n$.'}
The nit operators in equation~(\ref{2002-statepart-nitopgen}) can be combined to a single
measurement. The corresponding `context operator' $C=F_1 F_2 \cdots F_k$ can be obtained by
taking different prime numbers as diagonal entries of $F_1,\ldots , F_k$ (cf. examples below).

There exist $n^k!$  sets of nit operators,
which are obtained by forming a $(k \times n^k)$--matrix
\begin{equation}
\left(
\begin{array}{ccccccccccccccccccccc}
q_1&\ldots &q_1 &\ldots &q_n&\ldots &q_n\\
&&&\ldots &&&\\
q_1&\ldots &q_n &\ldots &q_1&\ldots &q_n
\end{array}
\right)
\label{2003-garda1}
\end{equation}
whose rows are the diagonal components of $F_1,\ldots,F_k$  from equation~
(\ref{2002-statepart-nitopgen}),
by permuting its columns, and finally by reinterpreting the rows as the diagonal entries of the new nit operators
$F_1',\ldots,F_k'$.
This formal procedure is equivalent to permuting (the labels of) the $n^k$ product states.
One consequence of the rearrangement
is the transition from nonentangled eigenstates of the single particle states to entangled eigenstates thereof
(see example below).
No straightforward meaning could be associated to the new nit operators in this general case.
Note that all partitions discussed so far are equally weighted and well balanced,
as all elements of them contain an equal number of states.

\subsection{Examples: two three--state particle cases and entanglement}

An example for the two three--state particle case has been enumerated in
Ref. \cite{svozil-2002-statepart-prl}.
Recall that, in the simplest case, the two nit operators can be constructed according to
the scheme in equation~(\ref{2002-statepart-nitopgen}) and represented by
\begin{equation}
\begin{array}{llllll}
F_1&=&\{\{1,2,3\},\{4,5,6\},\{7,8,9\}\}&\equiv& {\rm diag} (a,a,a,b,b,b,c,c,c),\\
F_2&=&\{\{1,4,7\},\{2,5,8\},\{3,6,9\}\}&\equiv& {\rm diag} (a,b,c,a,b,c,a,b,c).\\
\end{array}
\label{2002-statepart-ps3e}
\end{equation}
If, on the other hand, $F_2= {\rm diag} (d,e,f,d,e,f,d,e,f)$
and $a,b,c,d,e,f,$ are six different prime numbers,
then, due to the uniqueness of prime decompositions,
the two trit operators
can be combined to a single
{\em context} operator
\begin{equation}
C=F_1\cdot F_2=F_2\cdot F_1=
{\rm diag} (ad,ae,af,bd,be,bf,cd,ce,cf)
\label{2002-statepart-ps3pd}
\end{equation}
which acts on both particles.
As $C$ has nine different eigenvalues, it separates the nine product states
completely and at once.

Just as for the two states per particle case \cite{DonSvo01},
there exist $3^2!=9!=362880$ permutations of operators
which are all able to separate the nine states.
According to equation~(\ref{2003-garda1}),
they are obtained by forming a $(2\times 9)$--matrix
whose rows are the diagonal components of $F_1$ and $F_2$
from equation~(\ref{2002-statepart-ps3e})
and permuting all the columns.
The resulting new operators are also valid trit operators; i.e.,
for every one of the new pair of partitions
(i) the set theoretic intersection of single elements of the two partitions,
one element per partition, is a single product state,
and
(ii) the union of all these
intersections obtained by (i) is just the set of product states.
(For a proof recall that every permutation amounts to a relabelling the product states.)

\begin{figure}
\begin{center}
\tabcolsep 0 cm
\begin{tabular}{ccc}
 \includegraphics[width=5.1cm]{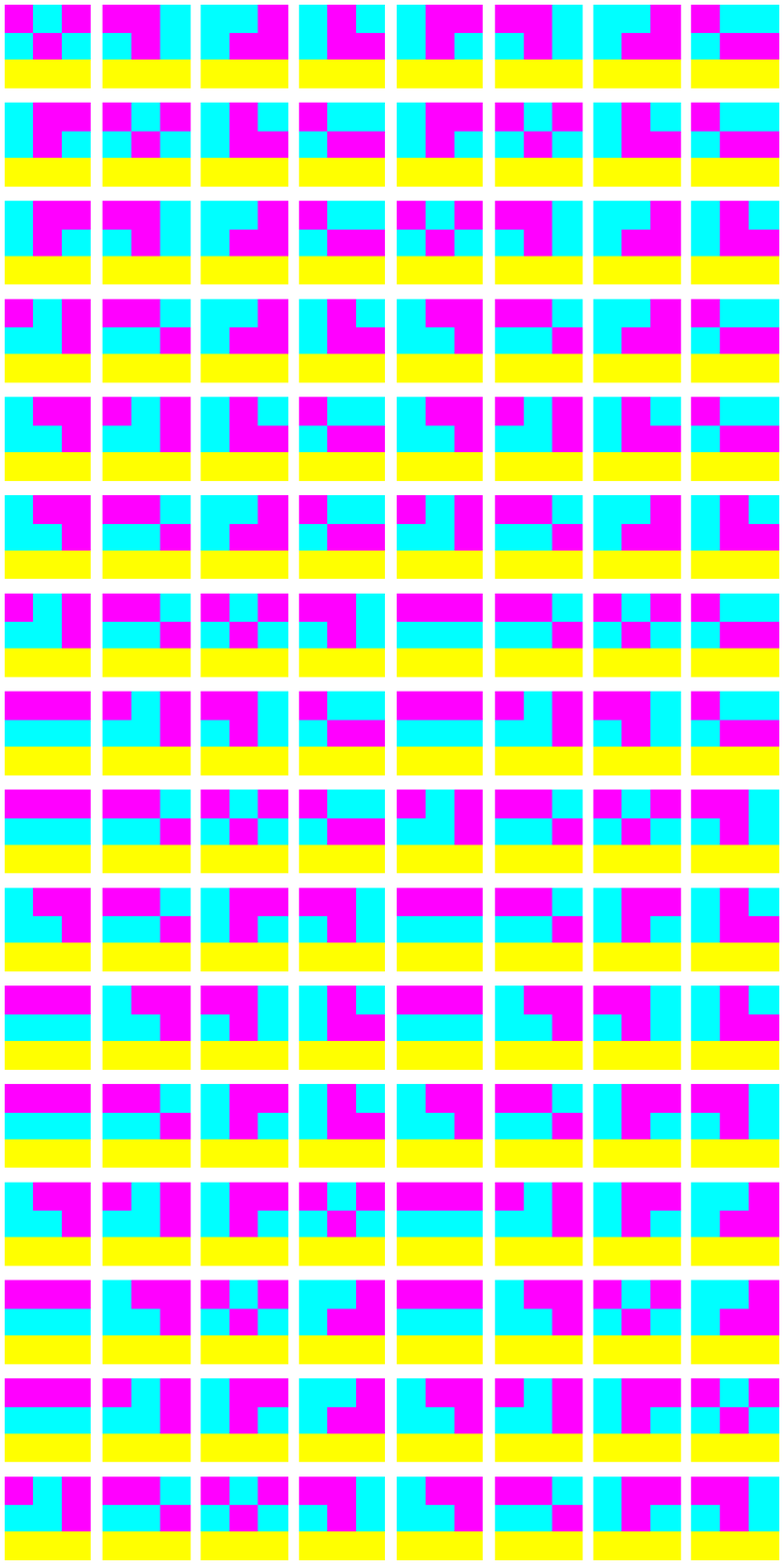} &
 \includegraphics[width=5.1cm]{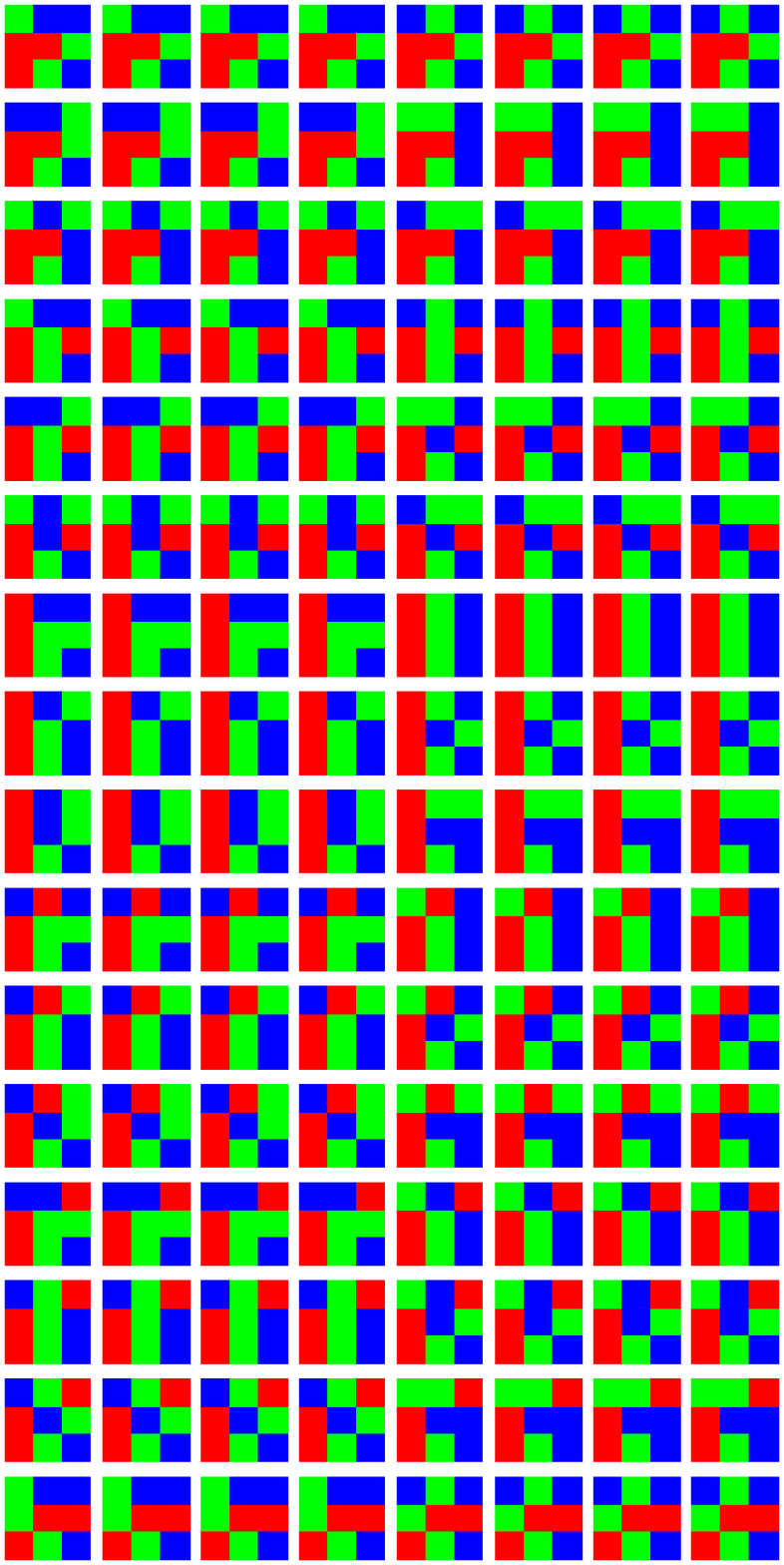} &
 \includegraphics[width=5.1cm]{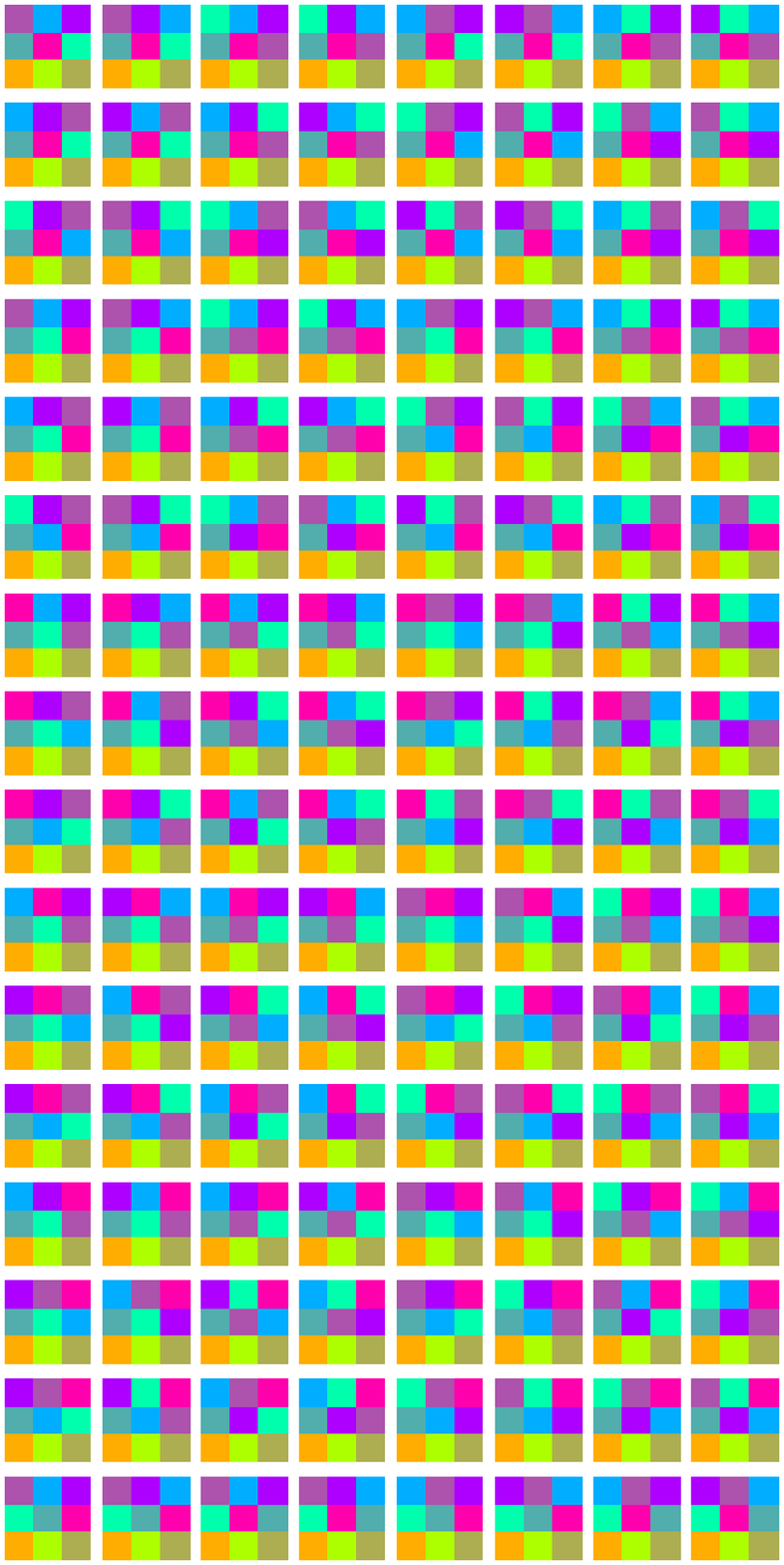}\\
trit 1&trit 2& trits 1\&2
\end{tabular}
\end{center}
\caption{Two trits yield a unique tessellation of the two particle product state space.
The first and second single particle states are drawn horizontally and
vertically, respectively.
Depicted are the first cases of equations~(\ref{2002-garda1-lb})---(\ref{2002-garda1-lf}).\label{2002-garda1-f1}}
\end{figure}

The complete set of $9!/(2\cdot 3!\cdot 3!)= 5040$
different two--trit sets can be evaluated numerically; i.e.,
in lexicographic order,
%<<DiscreteMath`
%
%p9=Permutations[{1,2,3,4,5,6,7,8,9}];
%
%trits[x1_,x2_,x3_,x4_,x5_,x6_,x7_,x8_,x9_]:=
%{Sort[{x1,x2,x3}],Sort[{x4,x5,x6}],Sort[{x7,x8,x9}],Sort[{x1,x4,x7}],Sort[{x2,x5,x8}],Sort[{x3,x6,x9}]};
%
%at=Union[Table[Sort[
%trits[p9[[i,1]],p9[[i,2]],p9[[i,3]],p9[[i,4]],p9[[i,5]],p9[[i,6]],p9[[i,7]],p9[[i,8]],p9[[i,9]]]
%],{i,9!}]];
\begin{eqnarray}
&\{\{\{\{1, 2, 3\}, \{4, 6, 8\}, \{5, 7, 9\}\}\times \{\{1, 4, 5\}, \{2, 6, 7\}, \{3, 8, 9\}\}\},  &\label{2002-garda1-lb}    \\
&\{\{\{1, 2, 3\}, \{4, 6, 9\}, \{5, 7, 8\}\}\times \{\{1, 4, 5\}, \{2, 6, 7\}, \{3, 8, 9\}\}\},  &\\
&\vdots         \nonumber                                                             &             \\
%&\{\{\{1,2,3\},\{4,5,6\},\{7,8,9\}\}\times \{\{\{1,4,7\},\{2,5,8\},\{3,6,9\}\}\} ,                  &     \\
&\{\{\{1,5,9\},\{2,6,7\},\{3,4,8\}\}\times \{\{\{1,6,8\},\{2,4,9\},\{3,5,7\}\}\} ,                  & \label{2003-garda-2}    \\
&\vdots         \nonumber                                                             &              \\
&\{\{\{1, 6,   9\}, \{2, 5, 7\}, \{3, 4, 8\}\}\times \{ \{1, 7, 8\}, \{2, 4, 9\}, \{3, 5, 6\}\}\},  &    \\
&\{\{\{1, 6,   9\}, \{2, 5, 8\}, \{3, 4, 7\}\}\times \{ \{1, 7, 8\}, \{2, 4, 9\}, \{3, 5, 6\}\}\}\}.& \label{2002-garda1-lf}
\end{eqnarray}
A graphical representation of the single particle state space tesselation
is depicted in figure~\ref{2002-garda1-f1}.

%\section{States}
%\subsection{Entanglement}

%The combinatoric methods introduced offer a wide variety of states to play with.
In general, the permutations transform nonentangled states into entangled ones.
Consider, for the sake of detail, the `(counter)diagonal' set of trits
listed in equation~
(\ref{2003-garda-2}),
which is induced by the permutation
whose cycle form
is (1)(2,5,6,7,3,9,8,4).
If the same two particle $(3\times 3)$ product state space representation is used
as in figure~\ref{2002-garda1-f1},
then the trits just correspond to the completed diagonals and counterdiagonals; i.e., if the
single particle states are labelled by
$a_1,b_1,c_1$
and
$a_2,b_2,c_2$,
respectively, then the new trit eigenstates
$\{
|\psi_1\rangle ,
|\psi_2\rangle ,
|\psi_3\rangle \}$
and $\{
|\psi_4\rangle ,
|\psi_5\rangle ,
|\psi_6\rangle  \}$ are
\begin{equation}
\begin{array}{llllll}
 |\psi_1\rangle &=& \frac{1}{\sqrt{3}} \left(|a_1a_2\rangle + |b_1b_2\rangle+ |c_1c_2\rangle \right) \equiv \frac{1}{\sqrt{3}} \left(1,0,0,0,1,0,0,0,1\right)^T ,\\
%|\psi_2\rangle &=& \frac{1}{\sqrt{3}} \left(|b_1a_2\rangle + |c_1b_2\rangle+ |a_1c_2\rangle \right) \equiv \frac{1}{\sqrt{3}} \left(0,0,1,1,0,0,0,1,0\right)^T,\\
%|\psi_3\rangle &=& \frac{1}{\sqrt{3}} \left(|c_1a_2\rangle + |a_1b_2\rangle+ |b_1c_2\rangle \right) \equiv \frac{1}{\sqrt{3}} \left(0,1,0,0,0,1,1,0,0\right)^T,\\
&\vdots &\\
%|\psi_4\rangle &=& \frac{1}{\sqrt{3}} \left(|a_1a_2\rangle + |c_1b_2\rangle+ |b_1c_2\rangle \right) \equiv \frac{1}{\sqrt{3}} \left(1,0,0,0,0,1,0,1,0\right)^T,\\
%|\psi_5\rangle &=& \frac{1}{\sqrt{3}} \left(|b_1a_2\rangle + |a_1b_2\rangle+ |c_1c_2\rangle \right) \equiv \frac{1}{\sqrt{3}} \left(0,1,0,1,0,0,0,0,1\right)^T,\\
|\psi_6\rangle &=& \frac{1}{\sqrt{3}} \left(|c_1a_2\rangle + |b_1b_2\rangle+ |a_1c_2\rangle \right) \equiv \frac{1}{\sqrt{3}} \left(0,0,1,0,1,0,1,0,0\right)^T
.
\end{array}
\label{2003-garda3}
\end{equation}
The associated trit operators are representable by
$F_1 = {\rm diag} (a,c,b,b,a,c,c,b,a)$ and
$F_2 = {\rm diag} (d,e,f,e,f,d,f,d,e)$, respectively;
with different $a=d$, $c=e$, and $b=f$;
or, alternatively, with mutually different numbers $a,b,c,d,e,f$.
With respect to the original single particle states, the trit eigenstates
(\ref{2003-garda3}) are
entangled.

\subsection{Inverse problems}
Consider the related dual or inverse problem:
suppose that a complete set of orthonormal states $S'$ is given;
what is the minimal set of comeasurable queries necessary to separate
any single one of these states from the other ones?
To answer this question, the unitary transformation $U$ connecting the
set of orthogonal states $S'$ with the standard orthonormal Cartesian basis $S$
can be used to transform the nit operators in equation~(\ref{2002-statepart-nitopgen})
into their appropriate form.

For two-state systems labelled by `$+$' and `$-$,'
the method can for instance be applied to
a set of orthonormal base states of eight dimensional Hilbert space which contains the
W--state introduced in \cite{zeil-97} and discussed in \cite{dvc-2000}.

 \begin{eqnarray}
 \begin{array}{lll}
 &&|\phi_1\rangle=  |+++\rangle  \\
 &&|\phi_2\rangle=\frac{1}{\sqrt{3}} \left( |++-\rangle+|+-+\rangle +|-++\rangle \right)  \\
 &&|\phi_3\rangle=\frac{1}{\sqrt{2}} \left( -|++-\rangle+|-++\rangle \right)  \\
 &&|\phi_4\rangle=\frac{1}{\sqrt{6}} \left( -|++-\rangle+2 |+-+\rangle -|-++\rangle \right)  \\
 &&|\phi_5\rangle=\frac{1}{\sqrt{3}} \left(  |+--\rangle+             |-+-\rangle +          |--+\rangle \right)  \\
 &&|\phi_6\rangle=\frac{1}{\sqrt{2}} \left( -|+--\rangle+|--+\rangle \right)  \\
 &&|\phi_7\rangle=\frac{1}{\sqrt{6}} \left( -|+--\rangle+ 2|-+-\rangle -|--+\rangle \right)  \\
 &&|\phi_8\rangle=  |---\rangle
 \end{array}
\label{2002-garda-f1}
 \end{eqnarray}

 Consider the unitary transformation $U^{\textrm{W}}$ given by
%   {{1, 0, 0, 0, 0, 0, 0, 0},
%    {0, 1/Sqrt[3], -1/Sqrt[2], -1/Sqrt[6], 0, 0, 0, 0},
%    {0, 1/Sqrt[3], 0, 2/Sqrt[6], 0, 0, 0, 0},
%    {0, 1/Sqrt[3], 1/Sqrt[2], -1/Sqrt[6], 0, 0, 0, 0},
%    {0, 0, 0, 0, 1/Sqrt[3], -1/Sqrt[2], -1/Sqrt[6], 0},
%    {0, 0, 0, 0, 1/Sqrt[3], 0,  2/Sqrt[6], 0},
%    {0, 0, 0, 0, 1/Sqrt[3], 1/Sqrt[2], -1/Sqrt[6], 0},
%    {0, 0,   0, 0, 0, 0, 0, 1}}
 \begin{equation}
 U^{\textrm{W}}= \left(
 \begin{array}{rrrrrrrr}
  1 & 0 & 0 & 0 & 0 & 0 & 0 & 0 \\
 0 & \frac{1}{{\sqrt{3}}} & -\frac{1}{{\sqrt{2}}} & - \frac{1}{{\sqrt{6}}} & 0 & 0 & 0 & 0\\
 0 & \frac{1}{{\sqrt{3}}} & 0 & \frac{2}{{\sqrt{6}}}                       & 0 & 0 & 0 & 0 \\
 0 & \frac{1}{{\sqrt{3}}} &  \frac{1}{{\sqrt{2}}} & -\frac{1}{{\sqrt{6}}} & 0 & 0 & 0 & 0\\
 0 & 0 & 0 & 0 &\frac{1}{{\sqrt{3}}} & -\frac{1}{{\sqrt{2}}} & - \frac{1}{{\sqrt{6}}}                                                                        & 0\\
 0 & 0 & 0 & 0 &\frac{1}{{\sqrt{3}}} & 0 & \frac{2}{{\sqrt{6}}}                                                                                             & 0\\
 0 & 0 & 0 & 0 &\frac{1}{{\sqrt{3}}} &  \frac{1}{{\sqrt{2}}} & -\frac{1}{{\sqrt{6}}}                                                                        & 0\\
 0 & 0 & 0 & 0 & 0 & 0 & 0 & 1 \\
 \end{array}  \right).
 \end{equation}
By construction, when
applied to the vectors of the standard orthonormal Cartesian basis,
$U^{\textrm{W}}$ yields the states
enumerated in equation~(\ref{2002-garda-f1}).
 The corresponding bit operators $F_1$, $F_2$, $F_3$ and the context operator $C$ are
\begin{equation}
 \begin{array}{lll}
 F_1&=&U^{\textrm{W}}\cdot \textrm{diag}\left(2,2,2,2,3,3,3,3\right)\cdot {U^{\textrm{W}}}^\dagger =\textrm{diag}\left(2,2,2,2,3,3,3,3\right)
 ,     \\
 F_2&=&U^{\textrm{W}}\cdot \textrm{diag}\left(5,5,7,7,5,5,7,7\right)\cdot {U^{\textrm{W}}}^\dagger
 ,       \\
 F_3&=&U^{\textrm{W}}\cdot \textrm{diag}\left(11,13,11,13,11,13,11,13\right)\cdot {U^{\textrm{W}}}^\dagger
 , \\
 C&=&F_1F_2F_3=U^{\textrm{W}}\cdot \textrm{diag}\left(110, 130, 154, 182, 165, 195, 231, 273\right)\cdot {U^{\textrm{W}}}^\dagger
 .
 \end{array}
 \end{equation}
Note that, if instead of
the prime numbers 2, 5, 11 and 3, 7, 13, we would have used 1 and 0, respectively,
projection operators would have resulted, but this strategy can
only be applied to the binary case \cite{DonSvo01}.

\section{Information of single quantum systems}

Having defined nits for the many--particle case,
let us now turn our attention to ome of the
mysterious and puzzling issues of quantum mechanics:
the postulated randomness of certain measurement outcomes
introduces an irreducible element of acausality.
Quantum randomness is accompanied by other principal limits
of operationalization and rational decidability,
such as complementarity and contextuality.
Encouraged by the conference agenda and
by many inspiring discussions with Professor Greenberger,
I shall raise a speculative and even controversial topic and
explore the randomness encountered in single and many--particle quantum systems
when there is a nit mismatch
between the states prepared and the states measured.

\subsection{Amazing single particle quantum systems}

Consider simple quantum mechanical preparation procedures, such as
the preparation of electrons in pure spin states along a particular direction
realizable by a Stern--Gerlach device.
Let us assume that we have prepared or `programmed' the electron spin to be in the `up' state
along our $z$--axis.
Then, by convincing ourselves that, when measured along $z$, the electron spin is always `up,'
we decide to ask the electron a `complementary' question, such as,
{\em ``what is the direction of spin along the $x$--axis perpendicular
to the  $z$--axis?'}
According to the quantum canon, in particular quantum complementarity,
the electron is totally incapable of `storing' precisely more
than one bit of information about its spin state in a single direction;
in particular it does not store
a second bit of information about its spin state
in any perpendicular direction thereof.
So, when interrogated about issues it was not at all prepared to answer,
it is at a complete loss of providing such information.

In this respect, the electron is like an input/output automaton
accepting only  sequences of strings consisting of the symbol
`$a$,' being confronted with the symbol `$b$.'
Indeed, to ridiculously overextend the `Copenhagen interpretation'
 to this automaton case, it would not make much sense
to push the word `$ab$' onto the automaton and watch its behaviour,
since such a behaviour property does not exist.
The query seems to be an absurd one in the sense of nonexistence of these properties.

Hence, quantum analogies with deterministic agents seem
to end when considering what happens in the case of absurd queries.
Deterministic agents are incapable of handling improper input,
on which they offer no answer at all.
The electron, on the contrary, seems to provide an answer,
albeit an irreducibly random one.
(In this case it behaves just like most Viennese when asked
about a location they do not know: they are too embarrassed to
confess their ignorance, so they will send
the questioner off into arbitrary directions;)

Thus, from the computational point of view, electrons are amazing little gadgets:
they are incapable of adding two plus two, let alone  universal computation;
yet in terms of algorithmic information theory \cite{chaitin:01,calude:02},
any humble electron seems to possess super--Turing computation powers.
To be more precise:  according to the
`creed'
canonized by some
`quantum council,'
the occurrence of certain individual quantum events are believed to be
totally unpredictable, unlawful, acausal,  and thus independent of past,
present and future states of the
system and of its surroundings, such as the measurement interface, in any algorithmically meaningful way
\footnote{We use the word `creed' here because this claim
cannot be operationalized, since it is impossible to devise a test
against all algorithmic laws.
The `quantum council' has been orchestrated by Bohr and  Heisenberg and adopted by
the majority of physicists; with irritating exceptions such as Schr\"odinger and Einstein.}.
As a consequence, with high probability,
algorithmically incompressible sequences can be generated from quantum coin tosses
\cite{svozil-qct,zeilinger:qct}.
%This is a very remarkable asset, indeed!
Summing up, in terms of spin, electrons seem to specialize in two antithetical tasks,
and in nothing else:
being prepared to issue a deterministic answer when asked a proper question; and
tossing more or less fair coins if asked improper questions.

\subsection{Quantum randomness through context translation}

We propose that the discrepancies of the seemingly inconsistent computational powers
of single quantum systems, such as the electron spin,
can be overcome by the assumption that it is not the electron which is the source of random data,
but the measurement apparatus and the environment of the measurement interface in general
which serves as a {\em `context translation'} of an improper question to a proper one,
thereby introducing noise.
The noise might originate from the many uncontrollable degrees of freedom of the measurement interface,
from the complex physical behaviour of the measurement apparatus, and from the observer in general.
The particular type of symmetries involved here seem to restrict the probabilities to Malus' law \cite{zeil-bruk-99}.

Let us consider possibilities to test and refute this  context translation by the interface.
One  operationalization would be the `cooling' of the interface
to produce a decrease of responsiveness of the measurement device.
It is to be expected that the ability to translate
the measurement context decreases as the temperature is lowered
and the many degrees of freedom which makes the measurement device quasi--classical are frozen.
This may also affect time resolution.
In this scenario,
in the extreme case of zero temperature,
the context translation might break down entirely,
and no discrimination between states could be given in the mismatch configuration:
The measurement device does not produce an answer to an improper question.

For a concrete example, consider a calcite crystal and polarization measurements of single photons
prepared in a linear polarization state along a single axis. If the  context translation hypothesis
were correct, the ability of an improperly adjusted calcite crystal to analyze the polarization direction of
photons would be diminished as it as well as the successive counters get cooler.
Close to zero temperature, for the mismatch configuration, there would not be any
polarization detection at all; the incoming photon would not get scattered and would remain at its original path.
As improbable as this scenario might appear, it is not totally unreasonable or inconsistent and should be
experimentally testable (e.g., see reference~\cite{Chaib-0953-8984-12-10-316} for a theory and
\cite{Herreros-Cedres:ks0116,kim-02} for experimental determinations of birefringence
in high--temperature ranges).

\section{Final remarks}

%Let us come back to the quasi-classical quantum doubles mentioned in the beginning.
%What kind of difference is there between a quantized system and the intrinsic perception
%of a deterministic agent incapable of answering all possible questions?
%Maybe there are less differences than expected.
%First and foremost, the Kochen-Specker theorem and the Boole-Bell conditions of possible classical experience
%rule out only the principle of omniscience;
%i.e.,  the assumption that all agents should be prepared to give answers to all conceivable properties,
%factual and counterfactual alike.
%(Scholasticism discussed this property under the name `infuturability.'
%Already very early Specker has related that ancient debate \cite{specker-60} to the algebraic
%logic of quantum propositions.)
%That is certainly not the case for finite automata or generalized urns.
%For these models it simply makes no sense to talk about certain classical properties at
%all---take the generalized urn
%example of the existence of blueness in the case if only yellow symbols are printed on the black background.
%

We have presented a formalization of nits for the many-particle case.
The present analysis is `top--down,' in that it is based on the standard formalism of
Hilbert space quantum mechanics.
From this point of view,
Zeilinger's foundational principle, which is intended as a `bottom--up'  principle,
is corroborated by the fact that,
quantum mechanically, with the nits properly defined via state partitions,
$k$ elementary systems can carry $k$ nits.
By this we mean that
$k$ mutually commuting
measurements of (joint or single particle)
observables with $n$ possible outcomes are necessary to determine the information
encoded in a quantum system completely.

We have also proposed a testable principle of context translation for the case
of a mismatch between state preparation and measurement.
With regards to this,
let us mention some amusing quasi--classical analogues.
Suppose, for instance, that you have just trained your refrigerator to tell you whether
or not it has enough milk for breakfast.
Then, if you asked the fridge whether there is enough butter in it,
maybe the best an intelligent program could do would be to guess the answer
on the basis of correlations of previous filling levels of milk and butter
and give a stochastic answer based on that sort of probabilities.
Yet the fridge might be at a complete loss if confronted with the question
whether or not there is enough oil in the car's engine.
If pressed hard,
it might toss a more or less fair coin and tell you some random answer,
if capable of doing so.

Instead of a refrigerator,
let us consider generalized urn models \cite{wright:pent,wright,svozil-2001-eua}
of the following form.
Suppose an urn is filled with black balls with coloured symbols on them, say blue and yellow.
Suppose further you have a couple of colour glasses of exactly the same colour.
Now if you draw a ball and look at it with such a coloured eyeglass,
you will only be able to perceive the symbols
in that particular colour, and not the other one(s). Conversely,
If you take another eyeglass,
you will see the symbols painted in that other colour.
A lot of fancy games can be played with generalized urn models; in particular complementarity games.
(Formally,
just as quantum mechanics,
their propositional structure is nonboolean;
i.e., nondistributive and thus nonclassical \cite{svozil-ql},
and turns out to be equivalent to automaton partition logic \cite{svozil-2001-eua}.
All finite quantum subalgebras are realized by these logics \cite[Section 3.5.3]{svozil-ql}.)
Consider a simple question: suppose that we are dealing with a two--colour model,
say blue and yellow, yet we pretend to look at the balls with a different colour, say green.
What will happen? Well, there are two cases, depending on the setup.
If our paints and filters are almost monospectral, we shall see
only black balls, because those balls were not prepared to give us `green' answers.
However, if the spectra of the paint and the filter are broadened as usual,
the original yellow and blue symbols will both appear green
(albeit darker and with less contrast than in the `true' colours).
If we expected a single unique symbol, we may be puzzled to see two symbols,
and we might wonder what the `message,' the `information' encoded in the ball is.
This occurs because of a mismatch between the original `information' prepared,
and the `information' requested by the observer.

The above models may be amusing anecdotes,
but are there any relevant connections with quantum physics?
And if so, are the analogies superfluous?
There is an obvious difference: The above examples are quasi--classical;
at any time the observer may switch from intrinsic to extrinsic mode by leaving
the incomplete knowledge standpoint inside the Cartesian prison \cite[Sect. 1.9]{descartes-meditation}.
For instance, an observer may just look up the oil level, or take off the coloured eyeglasses.
The difference between intrinsic and extrinsic standpoint is a system science issue
\cite{svozil-93,svozil-unev}.
In contrast, quantum mechanics does not offer such an escape from any sort of `Cartesian prison.'
It also seems to imply that there is nothing to escape to,
since,
by the various variants of the Kochen--Specker theorem
(e.g., \cite{ZirlSchl-65,kochen1,svozil-ql})
and bounds on classical probabilities by the Boole--Bell conditions of possible classical experience
(e.g., \cite{bell-87,2000-poly}),
there are certain properties whose mutual existence is inconsistent.
But maybe we are just too unimaginative to envision the many possible options
which we have
(cf. the context translation principle and \cite{pitowsky-82,clifton:99} for conceivable alternatives)?
Only future will tell, hopefully.

%\bibliography{svozil}
%\bibliographystyle{apsrev}
%\bibliographystyle{unsrt}
%\bibliographystyle{science}
%\bibliographystyle{numalg}
%\bibliographystyle{plain}

\end{document}